\documentclass[aps,twocolumn,showpacs,superscriptaddress,preprintnumbers,
amsmath,amssymb,floatfix,pre]{revtex4}  
\usepackage{graphicx}  
\usepackage{dcolumn}   
\usepackage{bm}        
\usepackage{amsmath}   
\usepackage{amsfonts}
\usepackage{amssymb}
\usepackage{natbib}
\begin{document}

\title{Time dependent elastic response to a local shear transformation in amorphous solids}
\author{F. Puosi}
\email{francesco.puosi@ujf-grenoble.fr}
\affiliation{Univ. Grenoble 1/CNRS, LIPhy UMR 5588, Grenoble, F-38041, France}

\author{J. Rottler}
\email{jrottler@phas.ubc.ca}
\affiliation{Department of Physics and Astronomy, The University of British Columbia,6224 Agricultural Road, Vancouver, British Columbia V6T 1Z4, Canada}

\author{J-L. Barrat}
\email{jean-louis.barrat@ujf-grenoble.fr}
\affiliation{Univ. Grenoble 1/CNRS, LIPhy UMR 5588, Grenoble, F-38041, France}
\affiliation{Institut Laue-Langevin, 6 rue Jules Horowitz, BP 156, F-38042 Grenoble, France}
\date{\today}

\begin{abstract}
The elastic response of a two-dimensional amorphous solid to induced local shear transformations, which mimic the elementary plastic events  occurring in deformed glasses, is investigated via Molecular Dynamics simulations. We show that for different spatial realizations of the transformation, despite relative fluctuations of order one, the long time equilibrium response averages out to the prediction of the Eshelby inclusion problem for a continuum elastic medium. We characterize the effects of the underlying dynamics on the propagation of the elastic signal. A crossover from a propagative transmission in the case of weakly-damped dynamics to a diffusive transmission for strong damping is evidenced. In the latter case, the full time dependent elastic response is in agreement with the theoretical prediction, obtained by solving the diffusion equation for the displacement field in an elastic medium.

\end{abstract}

\maketitle

\section{\label{sec:intro}Introduction}

In the last two decades, the understanding of plasticity in amorphous systems has greatly benefited from  numerical simulations (for recent reviews, see \cite{FalkEPJB10,revBarratLemaitre11,RodneyMSME11}). It is now well established that at low temperature the onset of plastic, irreversible deformation is due to the accumulation of elementary plastic events, consisting of localized, in space and time, atomic rearrangements involving only a few tens of atoms. This events were first identified by Argon \cite{Argon1979_1,Argon1979} and later described by Falk \textit{et al.} \cite{FalkPRE57} in terms of Shear Transformations (STs) or Shear Transformation Zones (STZ). Recent experiment in colloidal glasses supported this idea \cite{Schall07}. STs have been extensively studied in atomic scale simulations.  Athermal quasi-static simulations (AQS), consisting in applying quasi-static deformation to zero-temperature solids, have made it possible to identify  unambiguously single STs, allowing to study their spatial organization and size distribution \cite{MaloneyPRL93_b, MaloneyPRE74, TanguyEPJE06}. 

Unlike dislocations in crystals, STZs cannot be identified a priori. Therefore, the possibility to predict regions liable to plastic rearrangement has attracted considerable interest. Criteria have been proposed based on the observation of particle displacement fields \cite{LemaitrePRE07}, local elastic moduli \cite{dePabloPRL04,TsamadosPRE09} or {\textquotedblleft soft spots\textquotedblright} from low-frequency vibrational modes \cite{ManningPRL11}. On the other hand, the description of the consequences of a localized plastic event also received attention. Localized plastic events induce long-range deformation in the system: the stress that was maintained by the particles involved in the rearrangement is  released to the neighbors, which act as continuum elastic body. The perturbation field has a quadrupolar symmetry and a decay away from the source characteristic of the Eshelby inclusion model \cite{Eshelby57,MaloneyPRE74,PicardEPJE04,TanguyEPJE06}. The emergence of this behavior takes place within a finite time, corresponding to the propagation of the elastic signal in the system.  Rather surprisingly, a clear description of this mechanism is still missing.

Modeling the reaction of the elastic matrix to plastic rearrangements is a key element in several mesoscale approaches for the flow of amorphous solids \cite{BaretPRL02,PicardPRE05,MartensSM12,NicolasPRL13}. Long range effects are taken into account via elastic propagators having the four-fold quadrupolar symmetry, supported by experimental and numerical observation. However, all these models assume that the system response is instantaneous. This last point can clearly be improved by introducing a transmission mechanism with a finite speed. On these basis, it is apparent that the lack of a microscopic description of the elastic propagation after a ST is a strong limitation. 

The present paper addresses, using atomic scale simulations, the fundamental problem of the propagation of the elastic perturbation due to a ST in amorphous systems. Instead of looking for single plastic events in non-equilibrium simulations, we follow a different but equivalent approach, consisting of inducing artificial STs in a quiescent system. In order to investigate how inertia affects the response, different conditions of the underlying dynamics, from overdamped to underdamped, are considered. This is motivated by recent work by Salerno \textit{et al.} \cite{SalernoPRL12} where the role of inertia on the critical behavior of avalanches in strained amorphous solids is discussed.

The paper is organized as follows. Details about the model and  the procedure to simulate artificial shear transformations are given in Section \ref{sec:methods}. In Sec. \ref{sec:theory} we  first review the  Eshelby model for circular inclusion, then we develop, according to continuum elasticity theory, the full time dependent elastic response. The results of numerical simulations are discussed and compared to theoretical predictions in Sec.  \ref{sec:results}. The final Sec. \ref{sec:con} provides a short summary and discussion.

\section{Methods}\label{sec:methods}

\subsection{Model}

We consider a generic two-dimensional (2D) model of glass, consisting of a mixture of A and B particles, with $N_A=32500$ and $N_B=17500$, interacting via a Lennard-Jones potential 
  $ V_{\alpha\beta}(r)= 4 \epsilon_{\alpha \beta}\left[ \left( \frac{\sigma_{\alpha\beta}}{r} \right)^{12} - \left( \frac{\sigma_{\alpha\beta}}{r} \right)^{6}  \right ] $
with $\alpha,\beta=A,B$ and $r$ being the distance between two particles.  The parameters $\epsilon_{AA}$, $\sigma_{AA}$ and $m_A$ define the units of energy, length and mass; the unit of time is given by $\tau_0=\sigma_{AA}\sqrt{(m_A/\epsilon_{AA})}$. We set $\epsilon_{AA}=1.0 $, $\epsilon_{AB}=1.5 $, $\epsilon_{BB}=0.5 $, $\sigma_{AA}=1.0$, $\sigma_{AB}=0.8$ and $\sigma_{BB}=0.88$ and $m_A=m_B=1$. The potential is truncated at $r=r_c=2.5$ for computational convenience. The system sizes  $L_x=L_y=205$ are fixed and periodic boundary conditions are used. The equations of motion are integrated using the velocity Verlet algorithm with a time step $\delta t=0.005$. The temperature $T$ is controlled via a Langevin thermostat \cite{LangevinThermo}; the associated equations of motion are:
\begin{eqnarray}
\frac{d\mathbf{r}_i}{dt}&=&\frac{\mathbf{p}_i}{m}\\
\frac{d\mathbf{p}_i}{dt}&=&-\sum_{j\neq i}\frac{\partial V(\mathbf{r}_{ij})}{\partial \mathbf{r}_{ij}}-\frac{\mathbf{p}_i}{\tau}+\eta_i \label{eqn:lang2}
\end{eqnarray}
where $(\mathbf{p}_i,\mathbf{r}_i)$ are the momentum and the position of particle $i$, $-\frac{\mathbf{p}_i}{\tau}$ is a damping force and $\eta_i$ a random force obeying $\langle \eta_i(t) \eta_j(t') \rangle=(2k_BTm_i/\tau)\delta_{ij}\delta(t-t')$. This thermostat introduces a characteristic timescale $\tau$, related to the relaxation of temperature fluctuations. In the next sections, results for different values of the damping time $\tau$ are discussed.

The glassy states were prepared by quenching at constant volume equilibrated systems at $T=1$ to zero temperature with a fast rate $dT/dt=2\times 10^{-3}$. The shear and bulk modulus, $G_2=17$ and $K_2=98$,  have been measured with the method described in Ref. \onlinecite{MizunoPRE13}; the associated Poisson ratio is $\nu_2=0.70$ (the subscript is used to indicate two-dimensional quantities).

\subsection{Fictitious local shear transformations}

A local shear transformation is replicated by shearing  along $x$ and $y$ directions particles inside a circular region of radius $R$, that will be designated as shear transformation region (STR),  to distinguish it from authentic STZs. We fix the radius at $R=2.5$; this corresponds to about $n\simeq23$ particles inside a STR, which is consistent with the number of particles involved in a 2D shear transformation \cite{TanguyEPJE06}. The center of the STR defines the origin of our coordinate system $(x,y)$.
The coordinates of particles inside the STR are transformed according to:
\begin{equation} \label{eqn:ST}
\left\{ 
 \begin{array}{lr}
  x_i\rightarrow x'_i=x+\epsilon y \\
  y_i\rightarrow y'_i=y+\epsilon x 
 \end{array}
\right.
\end{equation}
where $\epsilon$ is the shear strain. The transformation is instantaneous and set the time origin. Particles inside the STR are frozen while the behavior of the surrounding one at later times is observed.

In order to reduce the noise, the response of the system is averaged over an iso-configurational ensemble with 10 trajectories. The angular bracket $\langle \rangle_{ic}$  will indicate the iso-configurational average, where particles start from the same positions but have different momenta. In addition, we introduce an average over disorder, i.e., over realizations of the ST in different positions of the system. In particular, the disorder average involves 48  different STRs. The combination of iso-configurational and disorder average will be indicated with simple  angular bracket $\langle \rangle$.
\begin{figure}[t]
\begin{center}
\includegraphics[width=5cm]{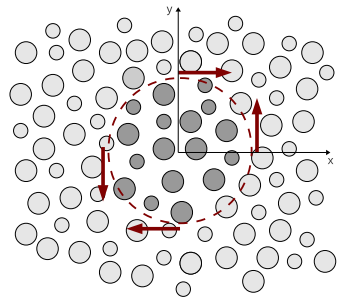}
\end{center}
\caption{ Sketch of a local shear transformation.  Frozen particles inside the shear transformation region STR  (dark grey particles),  a circular region of radius $a$, are instantaneously displaced along $x$ and $y$ directions according to the transformation defined in Eq. \ref{eqn:ST} }
\label{fig1}
\end{figure}

Now we examine briefly the shear strain $\epsilon$.  If $\epsilon$ is large, it could induce additional plastic events in other positions breaking down the elastic response. On the other hand, a small value for the strain could generate a too weak response to be detected. To set the best value for $\epsilon$ we operated as follows. We induce the shear transformation and let the system evolve for a time $\Delta t$, until everything has come to a new equilibrium state. Then we displace back to the original positions the particles inside the STR and let the system evolve again for a time $\Delta t$. The final configuration is compared to the initial one.  We observe that for $\epsilon=0.025$ the differences in the quenched energies are within the numerical precision. The associated relative displacement of particles are of the order of $1/10$ of the particle size, in agreement with the observations in spontaneous STs \cite{TanguyEPJE06}. In the next section, we show that  the perturbation in an elastic medium due to a ST is equivalent to that generated by two force dipoles of strength $f\simeq \epsilon G_2 R$. If one takes $R=2.5$ and $\epsilon=0.025$, then $ f \simeq 1$. In Ref. \onlinecite{LeonfortePRB70} it is shown that a source point force of order one is sufficiently small to ensure an elastic behavior in an amorphous elastic body.  We therefore adopt the  value  $\epsilon=0.025$ for the following investigation.
\begin{figure*}[t]
\begin{center}
\includegraphics[width=13cm]{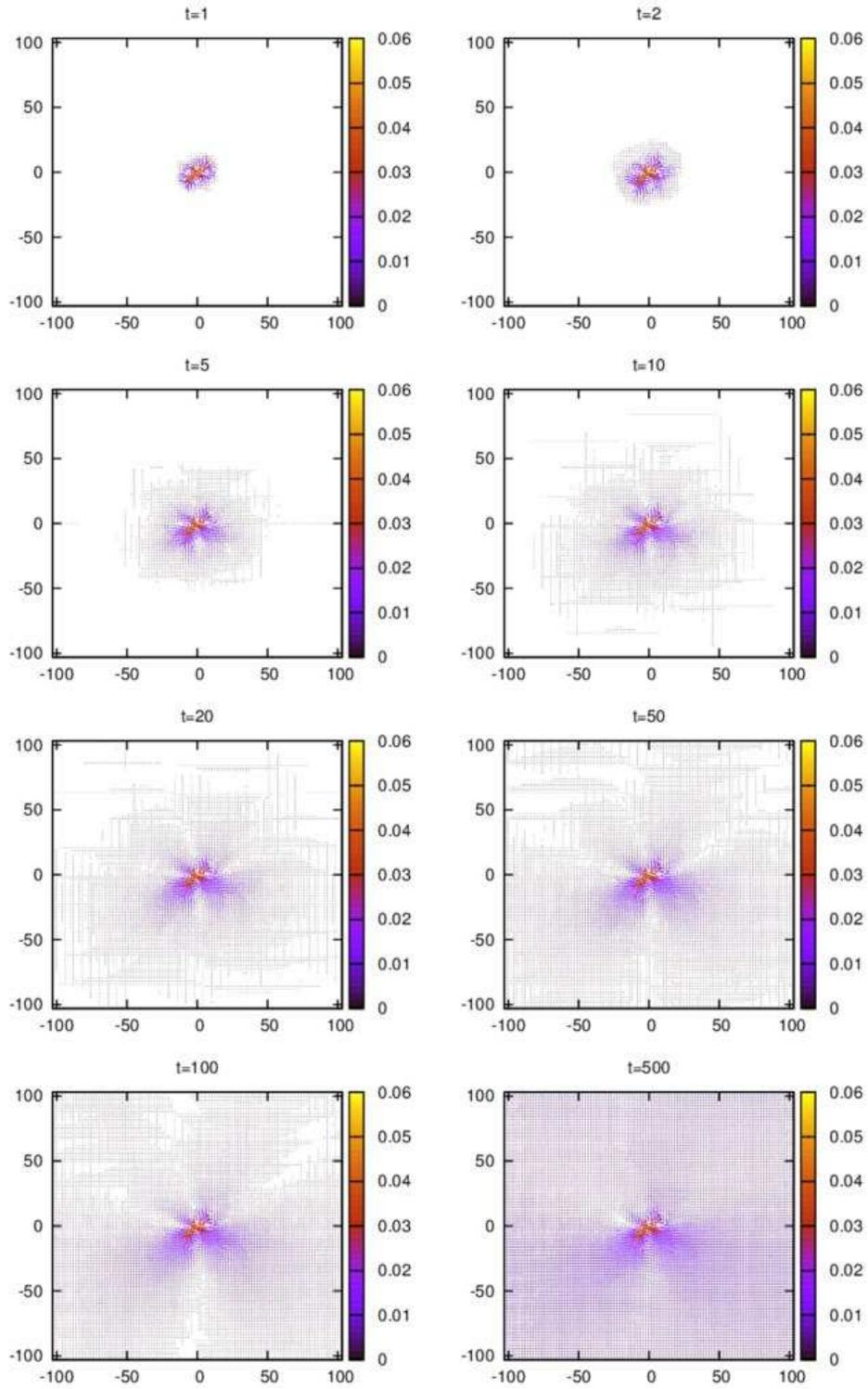}
\end{center}
\caption{ An example of the response to an induced local shear transformation. Snapshots  of the vector displacement field taken at different times which feature the propagation of the elastic signal in the system. }
\label{fig2}
\end{figure*}

\section{Time dependent displacement for a 2D circular inclusion}\label{sec:theory}

In this section we work out the analytical prediction for the displacement field due to a ST in the transient and equilibrium regime  according to classical elasticity theory.

We model the equilibrium situation as a 2D  Eshelby inclusion problem \cite{Eshelby57}.  We consider a  circular inclusion in a homogeneous elastic solid, that has been strained into an ellipse using an eigenstrain or stress-free strain $\epsilon^*_{\alpha\beta}=\epsilon^*(2\hat{n}_\alpha\hat{n}_\beta-\delta_{\alpha\beta})$, where $\epsilon^*$ is a scalar and $\hat{n}_\alpha$ a unit vector. The displacement field $u_\alpha(\mathbf{R})$ at a given point $\mathbf{R}$ in the elastic medium is the solution of the Lam\'{e}-Navier equation \cite{LandauEla}:
\begin{equation}\label{eqn:navelame}
 \left( \mu + \lambda \right)\frac{\partial^2 u_\beta }{\partial R_\alpha \partial R_\beta}+\mu \frac{\partial^2 u_\alpha}{\partial R_\beta \partial R_\beta}=0
\end{equation}
where $\mu$ and $\lambda$ are Lam\'{e} coefficients. Boundary conditions for Eq. \ref{eqn:navelame} are given by the expression of the field at the inclusion boundary, provided by Eshelby theory, and the requirement that the field vanishes for $r \rightarrow \infty$. The explicit solution \cite{ProcacciaPRE87} is then: 
\begin{widetext}
\begin{equation} \label{eqn:esh1}
	\mathbf{u}(\mathbf{r})=\frac{\epsilon^*}{4(1-\nu)}\left( \frac{a}{r} \right)^2 \left\{ \left[ 2(1-2\nu) + \left( \frac{a}{r} \right)^2 \right] \left [ 2\hat{\mathbf{n}}  \left( \hat{\mathbf{n}}\cdot\mathbf{r}\right) -\mathbf{r} \right]+ 2 \left[ 1-\left( \frac{a}{r} \right)^2 \right] \left[ \frac{2\left(\hat{\mathbf{n}}\cdot\mathbf{r}\right)^2}{r^2}-1 \right]\mathbf{r} \right\}
\end{equation}
\end{widetext}
where $a$ is the radius of the inclusion and $\nu$ the Poisson ratio.
We point out that Eq. \ref{eqn:esh1} is correct in three dimension in plane strain conditions and then also in two dimensions. 
If $\phi$ is the angle between the unit vector $\hat{\mathbf{n}}$ and the $x$ axis, Eq. \ref{eqn:esh1} in cartesian components become:
\begin{widetext}
	\begin{eqnarray}
		u_{x\infty}=  \frac{\epsilon^*}{4(1-\nu)}\left( \frac{a}{r} \right)^2 \left\{ \left[ 2(1-2\nu) + \left( \frac{a}{r} \right)^2 \right] \left( x\cos 2\phi+y\sin 2\phi \right)+2 x \left[ 1-\left( \frac{a}{r} \right)^2 \right] \frac{ \left( x^2-y^2 \right)\cos 2\phi+2xy\sin 2\phi }{r^2} \right\}  \label{eqn:uinfx}  \\
		u_{y\infty}= \frac{\epsilon^*}{4(1-\nu)}\left( \frac{a}{r} \right)^2 \left\{ \left[ 2(1-2\nu) + \left( \frac{a}{r} \right)^2 \right] \left( x\sin 2\phi-y\cos 2\phi \right)+2 y \left[ 1-\left( \frac{a}{r} \right)^2 \right] \frac{ \left( x^2-y^2 \right)\cos 2\phi+2xy\sin 2\phi }{r^2} \right\} \label{eqn:uinfy}
	\end{eqnarray}
\end{widetext}
We note that shearing simultaneously along $x$ and $y$ direction, as in the case of the shear transformation we considered, corresponds to $\phi=\pi/4$.

To derive the expressions for the displacement field in the transient regime we follow the approach of Ref. \onlinecite{LiuArxiv13}. First, we  switch to a pure two-dimensional description of the problem and we focus on the overdamped limit. The tensor equation for the diffusion of the vector displacement field can be written  as: 
\begin{equation}\label{eqn:tenseq}
	\Gamma \frac{\partial u_\alpha}{\partial t}=\mu_2 \frac{\partial^2 u_\alpha}{\partial R_\beta \partial R_\beta}  + \frac{\mu_2}{1-\nu_2} \frac{\partial^2 u_\beta}{\partial R_\alpha \partial R_\beta}
\end{equation}
where $\nu_2=\nu/(1-\nu)$. The left side of Eq. \ref{eqn:tenseq} represents the damping with a coefficient $\Gamma$, related to the time parameter $\tau$  in the Langevin equation, Eq. \ref{eqn:lang2}, via $\tau=\Gamma^{-1}$). From the right-hand side one can define:
\begin{eqnarray}
D_1 &=& \frac{2}{1-\nu_2} \frac{\mu_2}{\Gamma} \\
D_2 &= & \frac{\mu_2}{\Gamma}  
\end{eqnarray}
corresponding to the diffusion coefficients in the longitudinal and transverse directions respectively. 

To solve Eq. \ref{eqn:tenseq} for the response to a ST, we notice that, in the limit $a\rightarrow 0$, the perturbation displacement is equivalent to the one induced by a set of two orthogonal force dipoles with magnitude $a^2\mu\epsilon^*$, located at the origin \cite{PicardEPJE04}. The Green's tensor $G_{ijk}(\mathbf{r})$ relates the displacement $\mathbf{u}$ to a source term $\mathbf{P}$ via:
\begin{equation}
u_k(\mathbf{r})=\int d\mathbf{r'}G_{ijk}(\mathbf{r}-\mathbf{r'})P_{ij}(\mathbf{r'})
\end{equation}
The Green's tensor associated to Eq.  \ref{eqn:tenseq} is given by \cite{LiuArxiv13}:
\begin{widetext}
\begin{eqnarray}
G_{ijk}(\mathbf{r},t) &=&  -\frac{1}{\mu_2 r} \left\{  \left[  \left ( \frac{1-\nu_2}{2} + \frac{8D_2t}{r^2} \right) e^{-r^2/4D_1t} - \left ( 1 + \frac{8D_2t}{r^2} \right) e^{-r^2/4D_2t}  \right]  \frac{r_ir_jr_k}{r^3} \right. \nonumber \\*
 &-& \left. \frac{2D_2t}{r^2} \left[   e^{-r^2/4D_1t} - e^{-r^2/4D_2t} \right] \phi_{ijk}  + \delta_{ik}\frac{r_j}{r}  e^{-r^2/4D_2t} \right \}
\end{eqnarray}
\end{widetext}
with $\phi_{ijk}=\delta_{ij}\frac{r_k}{r}+\delta_{ik}\frac{r_j}{r}+\delta_{jk}\frac{r_i}{r}$. Explicitly calculating the response for a shear transformation, we obtain: 
\begin{widetext}
\begin{eqnarray}
u_x (x,y,t)&=& \frac{2\epsilon^*a^2}{ r}  \left\{  \left[  \left ( \frac{1-\nu_2}{2} + \frac{8D_2t}{r^2} \right) e^{-r^2/4D_1t} - \left ( 1 + \frac{8D_2t}{r^2} \right) e^{-r^2/4D_2t}  \right]  \frac{x^2y}{r^3} \right. \nonumber \\*
 &-& \left. \frac{2D_2t}{r^2} \left[   e^{-r^2/4D_1t} - e^{-r^2/4D_2t} \right] \frac{y}{r}+ \frac{1}{2}\frac{y}{r}  e^{-r^2/4D_2t} \right \} \label{eqn:uxt} \\
u_y (x,y,t)&=& \frac{2\epsilon^*a^2}{ r}  \left\{  \left[  \left ( \frac{1-\nu_2}{2} + \frac{8D_2t}{r^2} \right) e^{-r^2/4D_1t} - \left ( 1 + \frac{8D_2t}{r^2} \right) e^{-r^2/4D_2t}  \right]  \frac{xy^2}{r^3} \right. \nonumber \\*
 &-& \left. \frac{2D_2t}{r^2} \left[   e^{-r^2/4D_1t} - e^{-r^2/4D_2t} \right] \frac{x}{r}+ \frac{1}{2}\frac{x}{r}  e^{-r^2/4D_2t} \right \} \label{eqn:uyt} 
\end{eqnarray}
\end{widetext}
If we take the limit $t \rightarrow \infty$ in Eq. \ref{eqn:uxt} and Eq. \ref{eqn:uyt}, we obtain exactly the expression for stationary field, respectively Eq. \ref{eqn:uinfx} and Eq. \ref{eqn:uinfy}, in the limit $a\rightarrow0$ with the product $\epsilon^*a^2$ kept constant.

\section{\label{sec:results}Results and discussion}

\begin{figure}[h]
\begin{center}
\includegraphics[width=8cm]{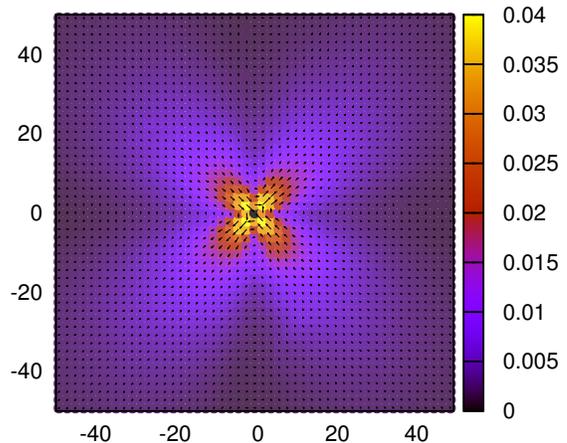}
\end{center}
\caption{ Long time mean displacement field  $\langle \mathbf{u}_\infty \rangle$ after a local shear transformation (in the origin) obtained averaging over realizations of the transformation in different regions of the system. The quadrupolar structure is equivalent to one observed in isolated plastic events occurring in sheared glasses.}
\label{fig3}
\end{figure}

A typical example of a fictitious shear event in a regime of intermediate damping $\tau=1$ is shown in Figure \ref{fig3} where we plot the time evolution of the  displacements $ \langle \mathbf{u}_i(t) \rangle_{ic}= \langle \mathbf{r}_i(t+t_0) -\mathbf{r}_i(t_0) \rangle_{ic} $ with the shear transformation taking place at time $t_0$.  The propagation of the elastic signal is apparent. At very short times only particles very close to the STR are affected by the transformation. Later the response propagates in the system and an increasing number of particles are displaced from the original position. At very long time, a new equilibrium state, different from the original one, is achieved; in the following we will refer to this configuration as the long time or stationary one. Realizations of shear transformations in different regions of the system result in extremely different transient and equilibrium patterns of the displacement field. This is a clear signature of the microscopic heterogeneity of the elastic properties which is a well known feature of glasses.

We focus on the mean response, i.e., the displacement field averaged over disorder. In Fig. \ref{fig3} the mean long time displacement field $\langle \mathbf{u}_\infty \rangle $ is shown for the intermediate damping case.
Not surprisingly, the fictitious shear transformation produces an elastic displacement field with quadrupolar symmetry. This agrees with the behavior observed in single localized plastic events occurring in amorphous systems under deformation. 

\begin{figure}[t]
\begin{center}
\includegraphics[width=8cm]{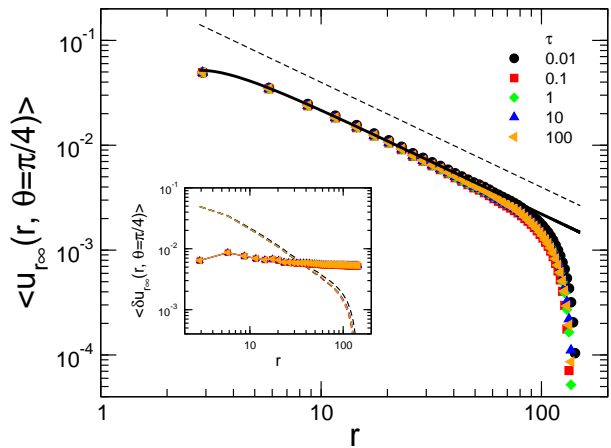}
\end{center}
\caption{ Main panel: symbols are the radial component of the long time displacement field $\langle u_{r\infty} (r,\theta) \rangle$, along the $\theta=\pi/4$ direction, for different values of the damping $\tau$. Both iso-configurational and disorder average are considered. Periodic boundary conditions are responsible for the field going to zero at the box boundaries. Full line is the prediction according to continuum elasticity theory, given by Eq. \ref{eqn:esh1}. Inset: fluctuations of the long time radial displacement $ \langle \delta u_{r\infty} (r,\theta) \rangle$ (symbols), compared to the mean displacement (lines). Relative fluctuations $\langle \delta u_{r\infty}  \rangle / \langle u_{r\infty} \rangle$ are found of order 1. }
\label{fig4}
\end{figure}

In order to test the prediction for the elastic response, because of symmetry reasons,  we move to a coordinate system $(r,\theta)$. In Figure \ref{fig4} we show the radial component of the long time field along the $\theta=\pi/4$  direction. Different values of the damping time $\tau$ are considered. 
Two point have to be made here. First, while the transient regime is expected to be strongly dependent on the damped dynamics (and this is the case as it will be shown later), the stationary solution is not: curves for different values of $\tau$, spanning from the very weak ($\tau=100$) to the strong damping  ($\tau=0.01$) regime, collapse. Second, data show a very good agreement with the prediction according to Eq. \ref{eqn:esh1}. We point out that no adjustable parameter was used in this comparison. The agreement is very good even for small distances from the sources, where the response is affected by the finite size of the STR.  The $1/r$-dependence of the radial component lasts until distances of the order of $L_{box}/4$, then the field drop to zero due to periodic boundary conditions. In the inset of Fig. \ref{fig4} we compare the average displacement with their respective fluctuation from sample to sample $\langle \delta u_\infty \rangle= \left( \langle u_\infty^2 \rangle - \langle u_\infty \rangle^2 \right)^{1/2}$. A different distance dependence between the two quantities is observed: while the mean radial displacement decays essentially analytically, due to the effect of boundary conditions, equilibrium fluctuations are almost flat (very weak exponential behavior for short distances $5 \lesssim r\lesssim 30$). Moreover, we note that fluctuations are of the order of the mean displacement, in agreement with the observation of Ref. \onlinecite{LeonfortePRB70}, where the response to a point source force was considered.

\begin{figure}[t]
\begin{center}
\includegraphics[width=8cm]{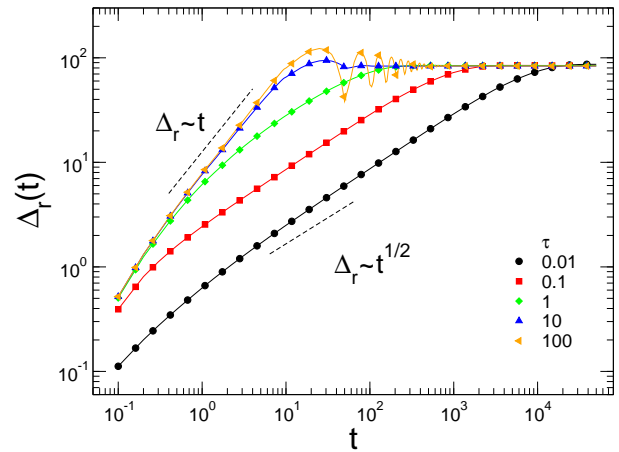}
\end{center}
\caption{ Time dependence of the total radial displacement  $\Delta_r(t)$, defined in Eq. \ref{eqn:DeltaR}, in different conditions of damping. The transmission of the elastic response to a shear transformation changes from propagative, $\Delta_r(t)\propto t$, at short time to diffusive, $\Delta_r(t) \propto t^{1/2}$,  at times longer than $\tau$ }
\label{fig5}
\end{figure}

Now we discuss the transient regime, in which the elastic signal propagates in the system. First we focus on the role played by the damping. We define the total radial displacement as:
\begin{equation}\label{eqn:DeltaR}
\Delta_r(t)= \iint  \vert \langle u_r(r,\theta,t) \rangle \vert r dr d\theta
\end{equation}
where the integration is performed over the full simulation box. $\Delta_r(t)$ gives a measure of the propagation of the elastic field. In fact, if one assumes that after a time $t$ the elastic signal has has traveled for a distance $R(t)$ and accordingly the radial displacement $u_r(r,\theta,t) $ is given by $u_r(r,\theta,t)=u_r^{cet}(r,\theta)\Theta\left( r- R(t) \right)$, where $u_r^{cet}(r,\theta)$ is the continuum elasticity expression, with a leading $1/r$ dependence, then one finds $\Delta_r(t)\propto R(t)$. In Fig. \ref{fig5} we show $\Delta_r(t)$ for the different conditions of damping. At short times, the transmission of the elastic response is propagative $\Delta_r(t)\propto t$, as can be clearly seen in the low-damped simulations ($\tau=10,100$). On the other hand, at times longer than $\tau$, diffusion controls the propagation of the displacement field and we have $\Delta_r(t) \propto t^{1/2}$.  

\begin{figure}[t]
\begin{center}
\includegraphics[width=7.5cm]{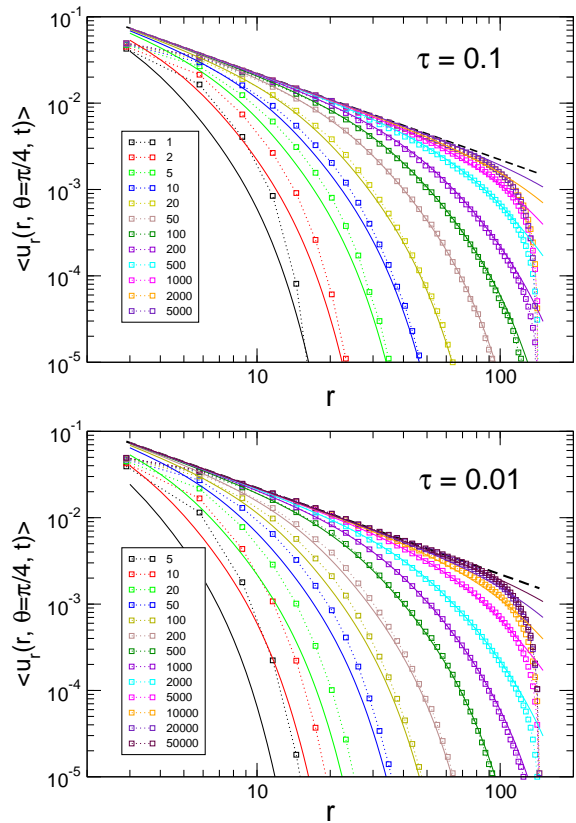}
\end{center}
\caption{ Open symbols: $\langle u_r (r,\theta,t)\rangle$, time dependent radial component of the displacement field along $\theta=\pi/4$ direction for highly-damped dynamics, $\tau=0.1$ (top) and $\tau=0.01$ (bottom). Both iso-configurational and disorder average are considered. Full lines: theoretical predictions according to Eq. \ref{eqn:uxt} and Eq. \ref{eqn:uyt}. }
\label{fig6}
\end{figure}

We are now in the position to compare simulation data with the full time dependent solution of the elastic response in the overdamped regime, namely Eq. \ref{eqn:uxt} and Eq. \ref{eqn:uyt}. Figure \ref{fig6} shows such a comparison.   We restrict to strongly-damped cases, $\tau=0.1$ and $\tau=0.01$, where we observe diffusive propagation. The agreement is surprisingly good and it improves further with increasing time.  Indeed, deviations can be seen for short times, where inertial effects are still present and the propagation is not clearly diffusive. 

\section{Concluding remarks}\label{sec:con}

We have investigated in atomic scale simulations the response of a standard 2D model of glass to a fictitious local shear transformation, which replicate the elementary plastic events observed in amorphous systems under shear deformation. Focusing on the displacement field, we fully characterized the propagation of the elastic signal for different conditions of the underlying dynamics.

First, we show that, despite large fluctuations (relative fluctuations of order one), the average (over different realizations of the ST) displacement agrees very well with the prediction of continuum elasticity theory  in both the stationary and transient regimes. A similar averaging behavior, restricted to the stationary regime, was observed for plastic T1 events in 2D simulations of foams under shear strain \cite{KablaPRL03}. 
 
Concerning the effect of inertia in the propagation of elastic response, our study may serve as an interpretation key for the results  of Ref. \onlinecite{SalernoPRL12} where the critical scaling of avalanches in quasistatic shear of disordered systems is discussed . They showed that $\Gamma_c=0.1$ ($\tau=10$) is a critical damping rate separating the overdamped (larger $\Gamma$) and underdamped (smaller $\Gamma$) regimes which are characterized by different scaling behavior.  Avalanches are due to the organization of individual ST, where  long-range elastic fields and the corresponding stress changes act as mechanical signals.   Now,  if we examine Fig. \ref{fig5} of the present work, we note that different propagation mechanisms dominate in the two previous limits,  propagative or ballistic for the inertial/underdamped  limit (large $\tau$) and diffusive for the overdamped one (small $\tau$),  with a crossover occurring for $\tau\sim 1-10$ ($\Gamma\sim 0.1-1$), in pretty good agreement with $\Gamma_c$. Therefore, we observe that the critical scaling behavior of avalanches, underdamped-like or overdamped-like, results from a particular propagation mechanism, propagative or diffusive respectively, for the elastic signals that trigger them. 

Lastly, we believe that the analysis of the present study can help in the improvement of mesoscale models for the flow of amorphous solids, in particular regarding the question of a finite time propagation and the effects of structural disorder, which represent the major drawbacks of several models \cite{PicardPRE05,MartensSM12,NicolasPRL13}. In this sense, finite element methods (FEMs), seem a promising approach since they can provide, by solving numerically the equation of the elastic equilibrium, the precise perturbation due to a ST, allowing in this way an immediate comparison with microscopic observations. This test will be the next step of the present work. 

\begin{acknowledgments}
We thank A. Nicolas and A.J. Liu for interesting discussions and H. Mizuno for providing the values of elastic constants.  The simulations were carried out using LAMMPS molecular dynamics software \cite{lammps} (\url{ http://lammps.sandia.gov}). JLB is supported by Institut Universitaire de France and by grant ERC-2011-ADG20110209.  JLB thanks the Consulate General of France in Vancouver and the Peter Wall Institute for Advanced Studies for support through the French Scholars Lecture Series.

\end{acknowledgments}

\bibliography{biblio_TDER}

\end{document}